%% file: psij.tex
\documentclass[conference]{IEEEtran}
\IEEEoverridecommandlockouts
\usepackage[dvipsnames]{xcolor}
\usepackage{cite}
\usepackage{amsmath,amssymb,amsfonts}
\usepackage{algorithmic}
\usepackage{graphicx}
\usepackage[hidelinks]{hyperref}
\usepackage{textcomp}
\usepackage{tikz}
\usepackage{graphicx}
\usepackage{booktabs}
\graphicspath{ {./images/} }

\usepackage{listings}

\definecolor{DarkGreen}{rgb}{0,0.2,0}

\def\BibTeX{{\rm B\kern-.05em{\sc i\kern-.025em b}\kern-.08em
    T\kern-.1667em\lower.7ex\hbox{E}\kern-.125emX}}
\newcommand{\seclabel}[1]{\label{sec:#1}}
\newcommand{\secref}[1]{Section~\ref{sec:#1}}
\newcommand{\figlabel}[1]{\label{fig:#1}}
\newcommand{\figref}[1]{Figure~\ref{fig:#1}}
\newcommand{\tablelabel}[1]{\label{table:#1}}
\newcommand{\tableref}[1]{Table~\ref{table:#1}}

\def\addvalue#1#2{\expandafter\gdef\csname note@colors@#1\endcsname{#2}}
\def\usevalue#1{\csname note@colors@#1\endcsname}
\def\setnotecolor#1#2{\addvalue{color-#1}{#2}}

\setnotecolor{AM}{Green}
\setnotecolor{MH}{Violet}
\setnotecolor{Kyle}{Red}
\setnotecolor{shantenu}{RubineRed}
\setnotecolor{Outline}{RubineRed}
\setnotecolor{mturilli}{Orange}
\setnotecolor{Ian}{blue}
\setnotecolor{AW}{SpringGreen}
\newcommand{\note}[2]{\par{\color{\usevalue{color-#1}} #1: #2}}
\newcommand{\inlinenote}[2]{~{\color{\usevalue{color-#1}} #1: #2}}
\newcommand{\resolvednote}[2]{}
\newcommand{\resolvedinlinenote}[2]{}
\newcommand{\discardednote}[2]{}

\unless\ifdefined\draft
    \renewcommand{\note}[2]{}
    \renewcommand{\inlinenote}[2]{}
\fi

\begin{document}
\bstctlcite{IEEEexample:BSTcontrol}

\title{PSI/J: A Portable Interface for Submitting, Monitoring, and Managing Jobs}

\author{
    \IEEEauthorblockN{
    Mihael Hategan-Marandiuc$^{1,4}$, 
    Andre Merzky$^{3}$, 
    Nicholson Collier$^{1,4}$,
    Ketan Maheshwari$^{6}$,
    Jonathan Ozik$^{1,4}$ \\
    Matteo Turilli$^{3,5}$,
    Andreas Wilke$^{1,4}$,
    Justin M. Wozniak$^{4}$,
    Kyle Chard$^{1,4}$, 
    Ian Foster$^{1,4}$, 
    Rafael Ferreira da Silva$^{6}$\\
    Shantenu Jha$^{3,5}$, 
    Daniel Laney$^{2}$
    }
    \IEEEauthorblockA{
        $^{1}$University of Chicago, Chicago, IL, USA
        $^{2}$Lawrence Livermore National;Laboratory, Livermore, CA, USA \\
        $^{3}$Brookhaven National Laboratory, Upton, NY, USA;
        $^{4}$Argonne National Laboratory, Lemont, IL, USA \\
        $^{5}$Rutgers, New Brunswick, NJ, USA;
        $^{6}$Oak Ridge National Laboratory, Oak Ridge, TN, USA
    }
}

\maketitle

\pagestyle{plain}

\begin{abstract}
It is generally desirable for high-performance computing (HPC) applications to be portable between HPC systems, for example to make use of more performant hardware, make effective use of allocations, and to co-locate compute jobs with large datasets. Unfortunately, moving scientific applications between HPC systems is challenging for various reasons, most notably that HPC systems have different HPC schedulers. 
We introduce PSI/J, a job management abstraction API intended to simplify the construction of software components and applications that are portable over various HPC scheduler implementations. We argue that such a system is both necessary and that no viable alternative currently exists. We analyze similar notable APIs and attempt to determine the factors that influenced their evolution and adoption by the HPC community. We base the design of PSI/J on that analysis. We describe how PSI/J has been integrated in three workflow systems and one application, and also show via experiments that PSI/J imposes minimal overhead.
\end{abstract}



\section{Introduction}\seclabel{intro}
\input{01-intro}

\section{Background and Related Work}\label{sec:background}
\input{02-background}

\section{Design}\seclabel{design}
\input{03-design}

\section{Deployment}\seclabel{deployment}
\input{04-deployment}

\section{Examples}\seclabel{examples}
\input{05-examples}

\section{Performance}\seclabel{performance}
\input{06-performance}

\section{Discussion and Next Steps}\seclabel{next}
\input{07-next}

 \section*{Acknowledgements}
 This research was supported by the Exascale Computing Project (17-SC-20-SC), a collaborative effort of the U.S. Department of Energy Office of Science and the National Nuclear Security Administration. This work was performed under the auspices of the U.S. Department of Energy by Lawrence Livermore National Laboratory under Contract DE-AC52-07NA27344 (LLNL-CONF-826133), Argonne National Laboratory under Contract DE-AC02-06CH11357, and Brookhaven National Laboratory under Contract DESC0012704. This research used resources of the OLCF at ORNL, which is supported by the Office of Science of the U.S. DOE under Contract No. DE-AC05-00OR22725. OSPREY work was supported by the National Science Foundation under Grant No. 2200234, the National Institutes of Health under grant R01DA055502, and the DOE Office of Science through the Bio-preparedness Research Virtual Environment (BRaVE) initiative. This manuscript has been authored in part by UT-Battelle, LLC, under contract DE-AC05-00OR22725 with the US Department of Energy (DOE). The publisher, by accepting the article for publication, acknowledges that the U.S. Government retains a non-exclusive, paid up, irrevocable, worldwide license to publish or reproduce the published form of the manuscript, or allow others to do so, for U.S. Government purposes. The DOE will provide public access to these results in accordance with the DOE Public Access Plan (http://energy.gov/downloads/doe-public-access-plan).

\Urlmuskip=0mu plus 1mu\relax
\bibliographystyle{IEEEtran}
\bibliography{references}

\end{document}

%% file: 01-intro.tex




A portable HPC application, intended to submit and manage jobs across multiple HPC systems, each with a potentially different \textit{local resource manager} (LRM), such as Slurm and PBS, must be able to generate job descriptions for each LRM type
and know how to interact with each LRM's
tools (e.g., qsub, qstat) or APIs. The semantics of these operations are, to a large extent, the same across the various LRMs and the differences are to be found in the details of how LRMs expose their functionality. Consequently, reasonable software architecture practices dictate that a portable HPC application be designed such that all the logic that is not specific to a given LRM be handled on an abstract level and the relevant translation to specific LRMs be done as low as possible in the application's component hierarchy. That is, under reasonable assumptions, every application intended to be portable across multiple LRMs must have a component that translates abstract job information into specific LRM scripts, commands, or API calls. For convenience, we will call such a component a Job Abstraction API (or, in short, JAAPI).

Without a common JAAPI, portable HPC applications have no choice but to implement their own custom solution. This leads to unnecessary redundancy, which could potentially be overlooked if the effort required to implement and maintain a JAAPI was negligible. However, evidence gathered from known past JAAPIs shows no indication of triviality in either design, implementation, or maintenance. On the contrary, it appears that the complexity of JAAPIs is frequently underestimated. This costly redundancy underlines the need for both reusable JAAPIs as well as an understanding of how their adoption can be facilitated and what prevents it.


The previous statements are abstract and do not do justice to just how prevalent this problem is. To provide some context, nearly every project that is incidental to our team's work (Balsam~\cite{Salim2019BalsamAS}, Maestro\cite{dinatale19maestro}, Parsl~\cite{babuji19parsl}, RADICAL-Pilot~\cite{merzky2021design}, Swift/T~\cite{wozniak2013swift}) 
has, separately, needed to design and implement a JAAPI, with little or no coordination but with very similar functionality, as shown in \tableref{apps}. And every JAAPI designed and implemented as part of such a project represents a hidden cost that must be funded but is rarely acknowledged as more than part of the normal cost of writing higher level HPC tools~\cite{wcs2022}.

\resolvednote{shantenu}{A fine point: some tools were preceded by the JAAPI that used them. Also, outside of ExaWorks, some large-projects like ATLAS and KEK adopted externally developed JAAPIs.}
\resolvednote{MH}{I should know. That's the case for Swift}
\resolvednote{MH}{I think this is fine. It's somewhat implicit in the general idea of how libraries are used.}

\resolvednote{MH}{Maybe we should actually make it clear on our web site and documents that we do want feedback on the spec, design, etc.}


A number of JAAPIs are known to have emerged over time. Of those that are still actively maintained or developed, none show evidence of widespread adoption. This is in direct contradiction with the clear need for a JAAPI. In this paper, we attempt to examine the causes of this contradiction and use our conclusions to provide insight and possible suggestions for facilitating the adoption of reusable JAAPIs as well as presenting recommendations for aiding in the survivability of JAAPIs. Informed by our analysis, we propose PSI/J: an open, language-agnostic, and minimalistic JAAPI that is meant to satisfy most of the needs of portable HPC applications while also providing pass-trough capability for more advanced usage scenarios.



\resolvednote{MH}{How this can also be used as a standalone tool to launch jobs in portable way; willnotfix}
\resolvednote{MH}{Should also emphasize community involvement somewhere}

We discuss past (and some present) similar efforts in an attempt to gain some insight into the reasons behind their limited availability and/or adoption in \secref{background}. With the wisdom gained in \secref{background}, we outline the design of the PSI/J API in \secref{design} as well as the relevant details of the Python binding~\cite{psijpython} and the testing infrastructure in \secref{deployment}. We proceed with a number of examples of applications that are using PSI/J in \secref{examples} and report performance numbers meant to illustrate the overhead that our PSI/J Python implementation imposes in \secref{performance}. We conclude with a discussion of the potential opportunities and obstacles that lie ahead in \secref{next}.


\begin{table*}
    \centering
    \renewcommand{\arraystretch}{1.3}
    \begin{tabular}{@{}c p{4.5cm} c c c c@{}}
     \toprule
     Application & Schedulers Supported  & Language & Remote \\
     
     \midrule
     Balsam~\cite{Salim2019BalsamAS} & Local, Cobalt, LSF, PBSPro, Slurm & Python & No \\


     Maestro~\cite{dinatale19maestro} & Local, Slurm, LSF, Flux & Python & No \\
     
     Parsl~\cite{babuji19parsl} & Cobalt, HTCondor, LSF, PBSPro, SGE, Slurm, Torque,  AWS, Azure, Google Cloud, Kubernetes  & Python &  Yes \\
    
     RADICAL-Pilot~\cite{merzky2021design} & Cobalt, Condor, CPI, LoadLeveler, LSF, PBS, PBSPro, SGE, Slurm, Torque & Python & Yes \\
          
     Swift/T~\cite{wozniak2013swift} & Cobalt, AWS, LSF, PBS, SGE, Slurm & Java, C, Tcl  & No \\
     \bottomrule
     \\
    \end{tabular}
    \caption{Select portable HPC applications and characteristics of their embedded JAAPIs.}\tablelabel{apps}
\end{table*}

\resolvednote{Kyle}{Table comparing the impls for different workflow systems. E.g., what scheudlers supported, what capabilities of scheduler are supported. Swift, Balsam, Flux, Maestro, Pegasus (HTCondor.. ), Radical/SAGA, DRMAA }

%% file: 02-background.tex



\begin{table*}
    \centering
    \renewcommand{\arraystretch}{1.3}
    \begin{tabular}{@{}c c p{4.5cm} c p{2,1cm} c p{2.5cm}@{}}
    \toprule
     JAAPI & Type & Schedulers Supported  & A/Synchronous & Languages & Remote & Last Updated \\

    \midrule
     DRMAA~\cite{troger07drmaa} & Spec. \& Impl. & SGE, HTCondor, PBS, LSF, GridWay, LoadLeveler, Slurm, Unicore & Async. & C, Perl, Python, Ruby, Java, Go, Erlang & Varies & Varies,  2017 (PBS), 2018 (Python) \\
     
     Globus GRAM~\cite{czajkowski98gram} & Impl. & PBS, SGE, LSF, Condor & Both & C, Python, Java & Yes & 2018 \\
     
     Unicore~\cite{romberg2002unicore} & Impl. & PBS, LoadLeveler, LSF, Slurm & Both & Java & Yes & 2023 \\
     
     HTCondor~\cite{thain05condor} & Impl. & PBS, LSF, SGE, AWS & Both & C, Python & Yes & 2023 \\

     SAGA~\cite{tom2006saga} & Spec. \& Impl. & SSH, HTCondor, PBS, SGE, Slurm, LSF, LoadLeveler, EC2 & Both & Python, C++, Java & Yes & 2023 (Python) \\

     Java CoG API & Impl. & PBS, Cobalt, HTCondor, SGE, Slurm, LSF, AWS, GCE, Globus GRAM, SSH & Both & Java & Yes & 2017 \\
     \bottomrule
     \\
    \end{tabular}
    \caption{A list of well known JAAPIs. Unicore, HTCondor, and RADICAL-SAGA are being actively maintained at the time of this writing.}\tablelabel{jaapis}
\end{table*}

In this section we describe both previous and current JAAPIs, discuss factors that influence the adoption and sustainability of JAAPI implementations, and analyze characteristics necessary to create a sustainable JAAPI. 

\subsection{Previous and current efforts}

The idea of a unified JAAPI is far from new. During the golden era of Grid Computing, there were a number of high profile JAAPIs. Perhaps one of the most notable 
was the Globus Toolkit~\cite{foster98globus} and, specifically, Globus GRAM~\cite{czajkowski98gram}. While now end-of-life, the Globus Toolkit is, directly and indirectly, a major source of inspiration for PSI/J.

Globus GRAM offers a set of services for submitting and monitoring jobs on diverse LRMs. Implemented as a set of services, and with a custom protocol and resource specification, GRAM offers a complete ecosystem for job management. GRAM is designed for Grid environments, and thus focuses on remote submission via integration of components for grid security, file transfer, and enforcement of authorization and access policies. GRAM is designed for deployment in multi-user environments and thus requires administrator privileges for deployment. 
At its peak, GRAM was deployed widely both in the US and internationally as an interface to different LRMs on many HPC clusters, including most of the leading supercomputers. 
Some of the ideas included in Globus continue
today in the cloud-hosted Globus service~\cite{chard14efficient} and in the  Globus Compute service~\cite{Chard2020funcX:}. 

After the first generation of Grid Computing tools, an effort emerged to build community consensus on Grid protocols and tools. This effort, the Open Grid Forum (OGF), was the birthplace of another major JAAPI: the Simple API for Grid Applications (SAGA)~\cite{tom2006saga}. An OGF standard, SAGA was implemented in C++~\cite{kaiser2006saga}, Java~\cite{russell2008vine} and, lastly, in Python via RADICAL-SAGA~\cite{merzky2015saga}.  RADICAL-SAGA is still maintained and used by RADICAL-Pilot~\cite{merzky2021design} and a few other distributed computing projects~\cite{merzky2015saga}
to interface with LRMs. 
SAGA and RADICAL-SAGA have directly inspired the design of PSI/J. OGF promoted a number of other notable efforts, of which we mention OGSA BES~\cite{foster2007ogsa} and JSDL~\cite{anjomshoaa2005job}. In the European Union, the UNICORE project~\cite{romberg2002unicore} provided a JAAPI in the form of the UNICORE Abstract Job Object.




At about the same time that SAGA was being designed, the Java CoG Kit~\cite{von2001java}, which started as a pure Java implementation of the Globus client suite, developed an ``abstraction API''~\cite{amin2004abstracting} for various job execution implementations, including Globus GRAM versions 2 and 3, SSH, LRMs, etc. This abstraction API would later be used by the Swift/K~\cite{zhao2007swift} workflow system both directly and indirectly as part of the Coaster pilot job system~\cite{hategan2011coasters}. Along with SAGA, it remains one of the main sources of inspiration for PSI/J.

Another product of the OGF is the Distributed Resource Management Application API~\cite{troger07drmaa} (DRMAA), which is an API that is meant to be exposed by HPC schedulers such that applications can submit and manage jobs through the DRMAA API rather than by interacting directly with the job scheduler. Applications are expected to link with the local DRMAA shared library and gain access to the local scheduler functionality. Implementations for DRMAA exist in C for Slurm, PBS/Torque, LSF. Wrappers for the C libraries can be found for Python, Perl, Ruby, Erlang, and Go. Since the peak days of the OGF, the HPC community seems to have moved forward and some of the C DRMAA implementations for certain LRMs have not been updated in more than 8 years. Nonetheless, DRMAA appears to be the most widely used JAAPI in existence.

\subsection{The JAAPI condition}

We now consider the factors that can influence the adoption and survivability of a JAAPI. We can classify relevant factors into two categories: \textit{intrinsic} to a particular design and/or implementation or \textit{extrinsic}. Intrinsic factors are factors that are almost entirely under the control of the team or organization enacting the design or implementation of a JAAPI, whereas extrinsic factors are a result of the circumstances under which the enactment takes place and for which the team or organization can at most produce mitigating strategies. This distinction between intrinsic and extrinsic factors can sometimes be blurry, but, in most cases, a dominant direction can be distinguished. In what follows, we will attempt to enumerate and discuss prominent factors. We caution the reader that this discussion is not meant as authoritative, but as a starting point for future discourse and, in the case of PSI/J, as the lens through which we perceive the suitability of a JAAPI solution. We will discuss extrinsic factors first, since they influence the intrinsic factors: complexity of the problem, funding structure, and the research nature of JAAPIs. We follow with the intrinsic factors: complexity of design, design inflexibility, and implementation quality.

\subsubsection{Problem complexity}

Designing and implementing a JAAPI is made complex by two characteristics of HPC systems: the heterogeneity of HPC sites and the stringent security associated with accessing most HPC sites. Even when different sites use the same LRM (e.g., Slurm), experience shows that there is significant variability in how LRMs are configured across sites. The assurance that a JAAPI functions properly on an HPC site rarely arises from it working on a similar site and is more often contingent on the JAAPI being tested on the specific HPC site. Consequently, ensuring that a JAAPI works in general requires that it be tested on a significant subset of the target set of HPC sites. But because of the stringent security policies associated with HPC sites, it is virtually impossible for a small development team to gain access to such a significant set of HPC sites.

We arrive at an odd intersection: nearly every portable HPC application that manages jobs needs to have a component that is nearly impossible to test adequately. Furthermore, the inability to test a component that sits at the base of the component hierarchy of an application has consequences for the entire application. While speculative, it does not stretch the imagination to picture why most HPC workflow systems, written with ad-hoc JAAPIs, fail to get traction outside the organizations in which they were created.

\subsubsection{Funding structure}

One of the major barriers to adoption of a JAAPI is the natural desire of adopters to have some confidence that the JAAPI will continue to be supported and maintained for the lifetime of the project that the JAAPI is meant to be integrated into. As it is undesirable to foresee the demise of one's own project, it follows that JAAPIs are expected to be supported for the foreseeable future in order for them to be a compelling target for adoption. Unfortunately, JAAPIs are mostly the products of research funding, which is, by nature, limited to a few years. Even if new funds are received, they must be justified in terms of an active research goal. This situation is problematic and leaves no room to the possibility of funding needed infrastructure. It is akin to requiring that highways be re-built with new materials or receive significant technological improvements every few years in order for them to continue to be in a usable state. Unfortunately, this results in higher overall costs, since lower quality JAAPIs now become a hidden cost of all portable HPC applications. Pushing the highway metaphor a bit further, it is like governments opting to fund multiple parallel roads for use by individual companies in order to avoid the costs of maintaining a single common road.

\subsubsection{Research nature of JAAPIs}

Traditionally, the largest customers of HPC resources have been scientific research projects. The biggest supercomputers are overwhelmingly associated with academic or research institutions. The ecosystem surrounding HPC resources and applications is, unsurprisingly, also tightly coupled with the research world. One of the main outputs in the research world are academic publications. Infrastructure software, such as JAAPIs, should have stability as one of their main goals. Stability implies changes that are limited to marginal improvements and the addressing of defects, changes that are rarely interesting enough to align with notable novelty that would result in research output. This leads to a conflict that, simultaneously, pushes JAAPIs towards stability and towards re-inventing themselves, which can add unnecessary complexity over time. Further, it encourages developers to focus on novelty, exploring new, yet untested, approaches necessary for publication.

\subsubsection{Complexity of design}

Complexity is perhaps one of the most visible aspects of a JAAPI solution. An overly complex design leads to difficulties in implementation, maintenance, and deployment. A complex design can arise in many ways. An often encountered scenario is that in which a large collaboration attempts to create an all-encompassing design. This leads to the inclusion of legitimate but rarely needed features that a compliant implementation is then forced to deliver. For example, the SAGA specification mandates that certain classes be implemented in versions that allow both synchronous and asynchronous invocation of their methods, even when such methods implement fast operations.  Complexity of design is not limited to a JAAPI specification and can apply equally to implementations.

\subsubsection{Design inflexibility}

Design inflexibility is the inability of a JAAPI design to adapt to changes in the software ecosystem used to enact implementations. It was typical of the JAAPIs that were products of the OGF and which included remote job submission to do so using an extensive stack of XML technologies which have since experienced a significant decline in adoption for new implementations. A subcategory of design inflexibility is programming language specificity, which explicitly or indirectly biases a design towards a specific programming language that may be subject to a decline in popularity.

\subsubsection{Implementation quality}

All JAAPI implementations are, in the end, software products. And, like other software products, they get consciously or subconsciously evaluated with respect to what potential users expect from a software product. Along with functional metrics, such as lack of errors, weight is also given to non-functional aspects such as code quality, presentation, documentation, ease of use, support, and so on. Ensuring that a well rounded software product is put forward requires team discipline that is often difficult to find in the research world where it is not uncommon for teams to consist of collaborations that are put together as part of a single funded project and where no clear organizational structure exists except PIs and not-PIs.

\subsection{Discussion}

We now turn towards an analysis of known JAAPIs as seen through the prism of the earlier considerations. We argue that a successful JAAPI must exhibit most of the desirable qualities mentioned previously while avoiding or suitably compensating for the difficulties. We, again, caution the reader: this analysis suffers heavily from survival bias as we only discuss JAAPIs that has managed to achieve sufficient prominence.

Perhaps the most successful JAAPI still in existence is DRMAA. The DRMAA specification is focused on a single problem and avoids many of the complexities associated with JAAPIs by only mandating local access to LRMs. While the DRMAA specification allows for implementations in multiple languages, it is primarily geared towards the C language, which is a form of design inflexibility. Libraries exist for all major LRMs, which, in large part, is a consequence of the DRMAA specification being written by representatives from LRM suppliers, some of which provide DRMAA libraries with LRM releases. The existence of the C libraries allow wrappers for other languages to be written without needing a full re-implementation. Nonetheless, the choice of most LRM specific DRMAA libraries to use proprietary APIs to communicate with the scheduler makes community support difficult, which is necessary without a continued source of funding or commitment from LRM suppliers. It also makes it challenging to maintain multiple LRM adapters as part of a single effort, which implies that some internal routines are duplicated across adapters and come with separate and often dissimilar characteristics. This appears to be a consequence of a design that fails to see beyond the implementation of DRMAA for a single LRM.

Still under active support, RADICAL-SAGA is an implementation of the SAGA specification. The SAGA specification is an extensive document that encompasses job management, file management, data stream management, remote procedure calls, and more. RADICAL-SAGA suffers from the rigidity imposed by having to adhere to this complex standard. For example, compliance requires a relatively complex implementation that limits portability and results in implementation quality issues, such as difficulty in tracing errors. In other words, the likelihood that a third party will implement SAGA and maintain the resulting library is significantly reduced. Indeed, the team maintaining RADICAL-SAGA, who are co-authors to this paper, have contributed significantly to the SAGA specification and maintain its Python implementation. Furthermore, RADICAL-SAGA's Grid Computing origins demanded a vast array of capabilities that are currently available through third party libraries. \resolvednote{shantenu}{whereas I'm sure the previous sentence is correct, I struggle a bit to understand it.}\resolvednote{mturilli}{better?}. \resolvednote{shantenu}{I would say other failures of RADICAL-SAGA consistent with the  desiderata of previous subsection, include: complexity of design and implementation quality (specifically poor error tracing and large dependency stack (contributing to a lack of portability))}\resolvednote{mturilli}{Added a sentence.}

Globus GRAM was an integral part of the Globus Toolkit. It allowed remote job submission when most people still accessed the Internet by dial-up. Implementing a remote-capable JAAPI required solving a number of security and networking problems for which no off-the-shelf solutions were readily available. Consequently, a great deal of complexity went into writing and maintaining it, which had ripple effects for maintenance costs. Additionally, in normal usage scenarios, the Globus GRAM server required a non-trivial amount of effort from HPC cluster administrators to be installed and supported. It was also initially closed-source, which contributed to the improbability of being supported by the community in the event of a funding shortage. Subsequent versions addressed the issue of community openness by releasing specifications through the OGF.

HTCondor started out as a cycle-stealing batch scheduling system, but has since grown into a system supporting the routing of jobs to other LRMs as part of a federated HTCondor pool. Various large Condor pools exist, including the Open Science Grid~\cite{osg}, but it is rare for leading HPC resources to be part of a HTCondor pool. HTCondor is a large and complex project with over 1.5k C++ source files and with many contributors. As a consequence, installing and configuring a local HTCondor pool that can act as a JAAPI is rarely the primary way in which portable HPC applications implement LRM abstraction.

The Java CoG Kit abstraction API was an implementation-only JAAPI that included support for job, file transfers, and remote filesystem access. While the Java GoG Kit abstraction API was, at first, a standalone product under the umbrella of the Java GoG Kit, it reached maturity almost exclusively as part of the Swift/K system at a time when Grid Computing was being slowly phased out. Combined with the fact that it was a Java-only implementation and that it suffered from implementation quality issues, such as poor documentation and support, it failed to gain widespread adoption. 

The UNICORE (Uniform Interface to Computing Resources) project developed an end-to-end middleware to enable users to execute many-task applications on Grid computing infrastructures. Evolved into a solution supporting general-purpose distributed computing, it requires root privileges on a cluster login nodes in order to deploy and run its server components on HPC platforms. Further, integration with each HPC cluster requires complex and specific customization. This imposes rigid constraints in terms of security, portability, and resource requirements and is an example of design complexity.

%% file: 03-design.tex



In this section we describe the goals that motivated PSI/J, the resulting specification, and the reference implementation.

\subsection{Design Goals}

\note{Kyle}{Perhaps the first aim should be portability/abstraction/generalizability, i.e., independent from a specific scheduler? }
\note{MH}{I think abstraction is implied through Job Absraction API and through the fact that the whole paper is about scheduler independent libraries}

The PSI/J API~\cite{psijspec} is designed as a minimal interface to submitting and monitoring jobs and their execution state. The need for minimalism is informed, in part, by the observations in \secref{background} that complexity is unlikely to lead to a successful JAAPI solution in the long term. That is coupled with the observation that many custom JAAPIs (e.g., Parsl~\cite{babuji19parsl}, Swift/K~\cite{zhao2007swift}, Swift/T~\cite{wozniak2013swift}, Balsam, RADICAL-Pilot) are focused almost exclusively on submitting and monitoring jobs with no further adornments. Therefore we believe that maintaining an orthogonal and therefore simple API is fundamental in facilitating both use and implementation.

The PSI/J API is also designed to allow scalable implementations where scalability is targeted both in the number of handled jobs and in the rate of job submission. Suggestions for implementers relevant to scalable implementations are provided in the specification document. Our reference Python PSI/J implementation makes full use of the scalability features of the API. Specifically, the API is asynchronous in order to support threadless use. The choice of asynchronous API is also motivated by the fact that one can easily transform an asynchronous API into a synchronous one, without incurring a performance penalty. However, the opposite is not true: adding an asynchronous layer on top of a synchronous API does not remove the one-to-one correspondence between threads and jobs at the synchronous API level.

\note{Kyle}{Extensible? Should be able to support all schedulers, irrespective of how they work (e.g., via API or batch submission script. Should be configurable to enable users to match against differnet options exposed by the specific schedulers.}

In PSI/J, the simplicity of the API is favored over that of the implementation if a reasonable implementation choice exists for a given design goal\cite{den2010price}. This is motivated by the fact that implementation complexity can be addressed with reusable solutions that multiple implementations can share, whereas API complexity translates into a pervasive overhead for users of the API. For example, bulk operations can, in certain cases, improve scalability of an implementation. Bulk operations are versions of API operations that act on multiple items as opposed to single ones. A bulk \emph{submit} method would take a list of jobs as arguments and submit them all in one call to the underlying LRM under the assumption that the alternative of making multiple calls to the LRM introduces a large combined overhead. However, a simple time windowing function can be used to aggregate individual job submissions and transparently route the resulting list of jobs to a bulk LRM submit call, if such a call is available. This would retain the simplicity of the API while also allowing the implementation to be efficient and is the favored choice. Certain caveats of this approach should, however, be noted. Time-based clustering of jobs is insufficient in determining jobs that are also related in their various properties and may be unsuitable in creating job arrays as supported by various LRMs. Such support may be added to PSI/J in the future.

A similar problem to bulk submission is that of bulk job status querying. Invoking \emph{qstat} commands individually for each job can result in poor scaling as the number of actively managed jobs increase. The PSI/J specification~\cite{psijspec} mandates that implementations query the status of jobs in bulk in order to avoid this issue.

Last but not least, the PSI/J specification is geared towards allowing implementations to live in user space as well as concurrently being installed as system libraries. This can push the burden of supporting a PSI/J implementation deployment away from system administrators and give users the flexibility to address deployment problems as needed. 

\subsection{The Structure of the API}

    \resolvednote{MH}{Think whether some pictures are suitable for the specification}
    \resolvednote{MH}{Andre \& Matteo: The pictures seem more related to the implementation, so consider moving them there}
    \resolvednote{MH}{These are meant to show how the spec informs an implementation rather than what the implementation necessarily is, although, sure, it is that because they work hand-in-hand.}

    The PSI/J specification is organized into three basic layers, depending on the level of functionality that it describes. The \emph{local layer} (see \figref{layer0}) defines the API needed to interact with job schedulers locally. That is, the location of the job scheduler is implicit and assumed to be on the same machine as the one on which the client application is running. The \emph{remote layer} (see \figref{layer1}) defines additional API elements needed to submit jobs to a remote scheduler, running on a different machine than the one on which client code is running. The \emph{nested layer} (Layer 2) is meant to add API elements needed to interact with pilot job implementations. At the time of the writing of this paper, only the local layer of the the PSI/J specification is available publicly and work is underway in drafting the remote layer. Nonetheless, the bulk of the specification rests with the local layer, with the other two layers containing only incremental updates needed to describe endpoints, credentials, and service configuration.

\newcommand{\cbox}[3]{
    \draw[rounded corners, fill=white] (#1) rectangle (#2)
        node[pos=0.5, align=center] {#3};
}

\tikzset{>=latex}
\tikzset{
    every picture/.style={line width=0.4pt}
}

\def\layerzero{
    \cbox{-0.2, 1.5}{8, 3.0}{PSI/J Core};

    \cbox{0, -1.2}{2, -0.2}{Slurm\\LRM}
    \cbox{2.2, -1.2}{4.2, -0.2}{PBS\\LRM}
    \cbox{4.4, -1.2}{6.6, -0.2}{LSF\\LRM}

    \draw[<->] (1, -0.2) -- (1, 1);
    \draw[<->] (3.2, -0.2) -- (3.2, 1);
    \draw[<->] (5.4, -0.2) -- (5.4, 1);

    \cbox{0, 1}{2, 2.0}{Slurm\\Executor}
    \cbox{2.2, 1}{4.2, 2.0}{PBS\\Executor}
    \cbox{4.4, 1}{6.6, 2.0}{LSF\\Executor}

    \draw (7.4, 1.1) node {\Large$\cdots$};
}
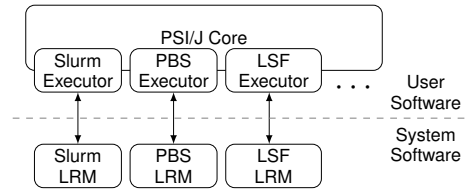
\begin{figure}[htb]
    \centering
    \scalebox{0.85}{
        \begin{tikzpicture}[font=\sffamily\footnotesize, scale=0.68]
            \layerzero
            \draw [ultra thin, gray, dashed] (-0.5, 0.4) -- (10, 0.4);
            \node[align=center] at (9, -0.2) {System\\Software};
            \node[align=center] at (9, 1) {User\\Software};
        \end{tikzpicture}
    }
    \caption{Illustration of the local layer of PSI/J.}
    \figlabel{layer0}
\end{figure}

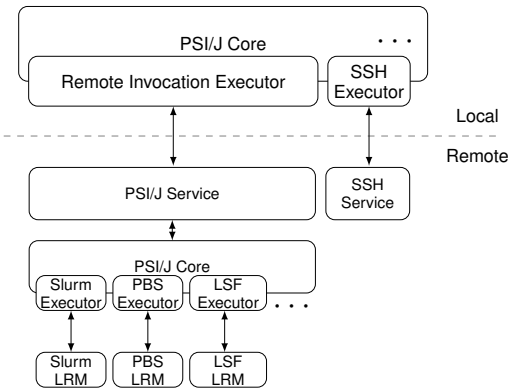
\begin{figure}[htb]
    \centering
    \scalebox{0.85}{
        \begin{tikzpicture}[font=\sffamily\footnotesize, scale=0.78]
                \begin{scope}[shift={(0.15, -3.8)}, scale=0.7, font=\sffamily\scriptsize]
                    \cbox{-0.2, 3.6}{8, 5.1}{PSI/J Service}
                    \cbox{8.3, 3.6}{10.7, 5.1}{SSH\\Service}
                    \layerzero
                    \draw[<->] (3.9, 3.6) -- (3.9, 3.0);
                \end{scope}
    
                \draw [ultra thin, gray, dashed] (-0.5, 0.4) -- (9.5, 0.4);
    
                \cbox{-0.2, 1.5}{8, 3.0}{PSI/J Core};
                \cbox{0, 1}{5.8, 2.0}{Remote Invocation Executor};
                \cbox{6.0, 1}{7.65, 2.0}{SSH\\Executor};
    
                \draw[<->] (2.9, -0.2) -- (2.9, 1);
                \draw[<->] (6.825, -0.2) -- (6.825, 1);
    
                \draw (7.4, 2.3) node {\Large$\cdots$};
    
                \node[align=center] at (9, 0) {Remote};
                \node[align=center] at (9, 0.8) {Local};
    
        \end{tikzpicture}
    }
    \caption{Intended usage scenario for the remote layer of PSI/J.}\resolvednote{shantenu}{be sure to remove ``possible'' from caption once settled/converged.}\resolvednote{MH}{It's "possible" as in "we don't implement this yet in the reference python implementation, but the spec is geared towards allowing such an implementation". It's not "possible" as in "diagram is not finalized". But maybe better to rename anyway.}
    \figlabel{layer1}
\end{figure}

The PSI/J layers are related by an inclusive relationship, in the sense that a nested layer implementation is expected to also contain all the API elements of the remote and local layers. However, one can also identify a functional relationship between layers. For example, the remote layer functionality can be implemented by adding remote invocation capabilities (see \figref{layer1}) to a local layer implementation, whereas a nested layer implementation requires a mechanism to submit the pilot jobs to a HPC scheduler, which can be achieved with either a local or remote layer. The remote layer is also suitable for interfacing with libraries that natively implement remote job execution, such as SSH.

In the remote layer, multiple remote schedulers on multiple HPC schedulers are meant to be accessible concurrently, from the same client process. Similarly, in the local layer, it can be desirable to be able to submit simple test jobs that can be run locally, using a forked process. To support this scenario, PSI/J-Python adopts a multiple-dispatch mechanism, in which bindings to underlying job execution mechanisms can coexist and be used concurrently. Such bindings are called ``executors''. In this sense, it differs fundamentally from most DRMAA implementations, which require that the client executable be dynamically or statically linked in order to switch to an alternate LRM.

\subsection{Implementations}

As previously stated, the implementations of PSI/J are meant to be usable as user space components. This is intended to overcome issues that stem from delays in updates of system software on HPC clusters. Cluster systems software, especially on HPC systems that are part of larger organizations, are known to go trough lengthy processes of approvals for updates on software packages that are deployed system-wide. This can result in large delays between the reporting of an issue with such software and the availability of a fix on a given HPC cluster.

We provide a reference Python implementation of PSI/J. The Python language was chosen due to its widespread use among higher level tools within the scientific community. However, in order to facilitate portability, most LRM specific functionality is described as textual templates using a templating library with binding in a large number of languages. The PSI/J specification and implementations can be logically split into two main parts: the \textit{core classes}, which are independent of the underlying execution mechanism (e.g., LRM), and the \textit{executors and launchers}, which implement LRM and cluster specific functionality. Executors are entities that know how to communicate with specific LRMs, whereas launchers describe the command used to launch multi-node jobs once the job resources are allocated. The core classes are used to build and manage jobs on an abstract level, independent of the underlying executor. Once a job, with all the necessary information, is built, one obtains an instance of an executor and uses it to submit the job. We show a simple example of this process for the Python binding of PSI/J in \figref{code-example}.

\vspace*{-0.1in}
\begin{figure}[htb]
    \centering
    \begin{lstlisting}[language=python]
 job = Job(
   spec=JobSpec(
     executable='/opt/cps/bin/NOARCH.x',
     arguments=['-qmp-geom', '8', '4', '4', '4',
                'do_arg.vml', 'evo_arg.vml',
                'eig_arg.vml', '0.00', 'Overlap'],
     stdout_path=cwd + '/eig.out',
     stderr_path=cwd + '/eig.err',
     resources=ResourceSpecV1(process_count=512),
     launcher='srun'
   )
 )
 ex = JobExecutor.get_instance('slurm')
 ex.submit(job)
    \end{lstlisting}
    \caption{Simple example for submitting a job in PSI/J Python with boilerplate removed.}
    \figlabel{code-example}
\end{figure}


The Python implementation of PSI/J uses a dynamic plugin discovery mechanism which allows a PSI/J core to detect executor and launcher implementations that are installed in different places from the PSI/J core. This allows for various scenarios, such as a stable system provided core using a user customized executor implementation or a user installed core using a system provided executor implementation (a scenario somewhat similar to that of DRMAA). Executors for Slurm, PBSPro, LSF, Cobalt, and Flux~\cite{flux} are provided with the current version of the reference implementation. A ``local'' executor that runs jobs using a simple fork mechanism is also provided. Launcher implementations are provided for all LRM specific launchers (\emph{srun}, \emph{aprun}, \emph{jsrun}, etc.) as well as for generic launchers, such as \emph{mpirun}. The dynamic plugin system allows PSI/J Python to be extensible with new executors and launchers without necessarily requiring that the additional executors or launchers be part of the PSI/J Python code base. 

We have opted for the executors provided by the reference PSI/J Python implementation to use publicly available LRM interfaces, which, in most cases, consist of well known commands, such as \textit{qsub} and \textit{qstat}. This is a deliberate choice that assumes that established public LRM interfaces are less likely to change and lead to incompatibilities than proprietary ones. This is in contrast to many DRMAA implementations which use proprietary APIs to communicate with LRMs. The choice made by the PSI/J Python reference implementation does not, however, preclude one from writing executors using proprietary APIs.

\resolvednote{AM}{RP is not a LRM really, I actually would opt for leaving it out in the list above.  Not sure how well that goes politically though...}\resolvedinlinenote{MH}{Removed reference to RP here}
\resolvednote{MH}{TODO: tie this up with a local conclusion; meh, I'm out of ideas}

%% file: 04-deployment.tex

One of the major problems that PSI/J is meant to address is the difficulty involved in adequately testing JAAPIs on a widespread set of HPC resources. We can assert that it is virtually impossible for a single research team to gain and maintain access to a sufficient number of resources in order to cover a satisfactory range of configurations. Traditionally, testing on resources that are not accessible to the JAAPI developers tends to be done by their users in the process of using the higher level system that employs the JAAPI. We propose to improve on this scenario by allowing any user with access to a HPC cluster (or any computing resource), when authorized to do so, to set up daily runs of a PSI/J implementation test suite and automatically report the results to a centralized test results aggregation site. The aggregation site can then be consulted by PSI/J developers who can react to issues detected on various systems.

The Python implementation of PSI/J allows its test suite to be configured to run daily on a given machine, typically using the standard \emph{Cron} tool, and upload detailed results of the tests to a central location. By default, the system is set up to test the main branch as well as branches belonging to GitHub pull requests that are made by the core team. While not currently implemented, we envision adding a tagging mechanism that allows the test suite to also be run on external pull requests, but only if tagged by a member of the core team. This allows both continuous testing of the main branch of the Python implementation of PSI/J, as well as that of relevant proposed changes. A \emph{minimal uploads} mode, which strips all non numeric and non boolean information from the results but retains the detailed information locally, is provided for use on sensitive and/or classified systems. When errors are detected on such a system and if desired, PSI/J developers have an option of emailing the test maintainers to ask for detailed information which can then be manually inspected and edited for sensitive information by the test maintainers.

\begin{figure}[htb]
    \centering
    \includegraphics[scale=0.32]{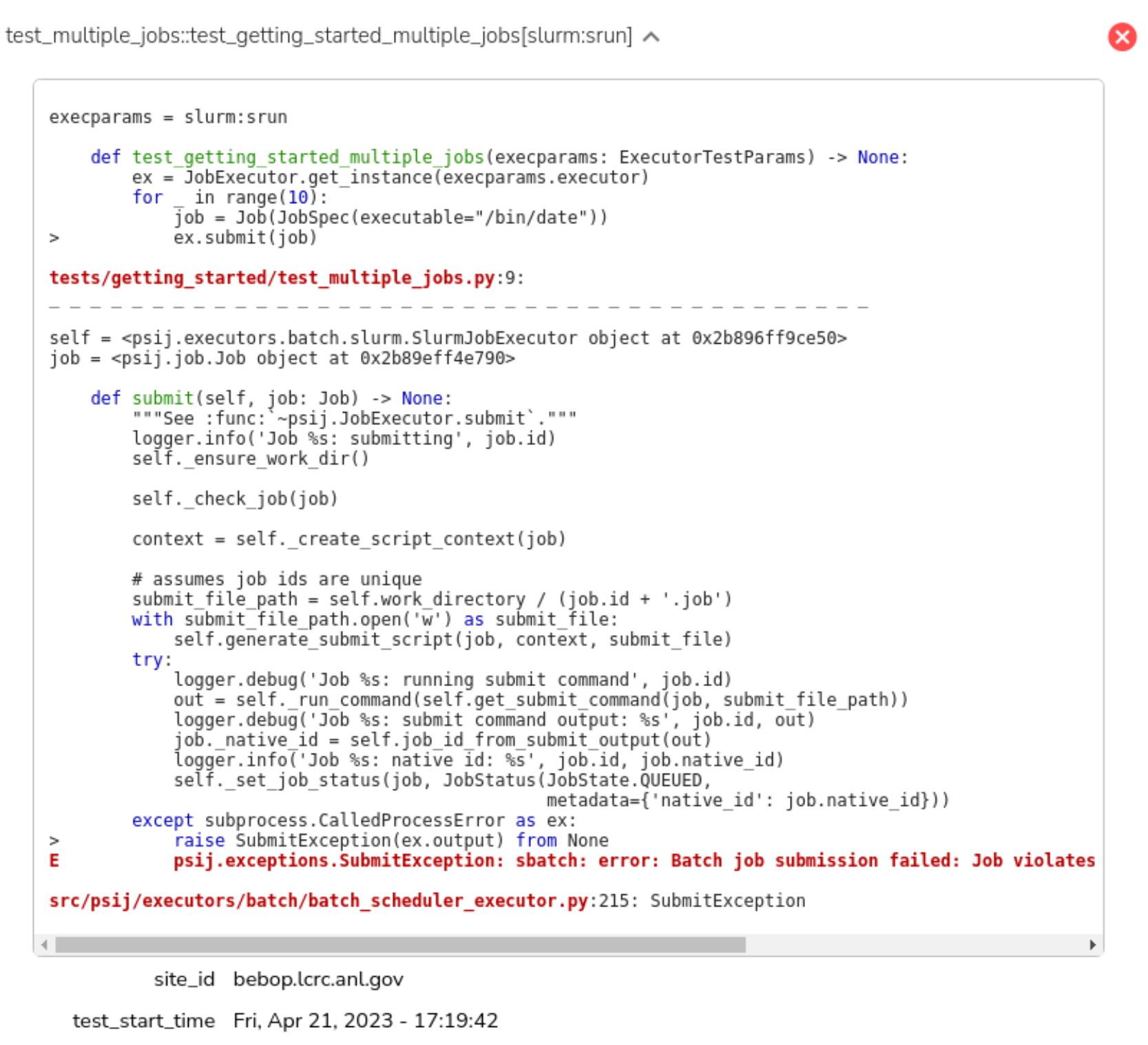}
    \caption{Screenshot of the test results view of the PSI/J testing dashboard showing a failed test and the Pytest report.}
    \figlabel{dashboard-test}
\end{figure}

The results of the tests are presented through a dashboard, which can display the data in multiple ways. One of the display modes shows aggregate results for each site over the past few days. Maintainers of PSI/J implementations can quickly identify sites on which one or more tests are failing. One can then select a particular site and inspect the actual \textit{pytest} output, and example of which is shown in \figref{dashboard-test}. This effectively allows maintainers and developers of PSI/J to run tests and get detailed results on resources that they would not normally have access to. It also unburdens them from setting up, maintaining, and prioritizing test sites while allowing users to gain confidence that PSI/J implementations function correctly on their sites.

%% file: 05-examples.tex

In this section we describe the use of PSI/J by three workflow programming frameworks, Parsl, RADICAL-Pilot, and Swift/T, and one application framework, OSPREY.

\subsection{Parsl}

Parsl is a parallel programming library for Python that allows users to write programs in Python that orchestrate parallel execution of Python functions and other external application. Parsl exposes a straightforward programming model in which developers identify opportunities for concurrent execution by adding decorators
to Python functions, called Parsl \textit{apps}. Any call to a Parsl app creates a new task that executes concurrently with the main program and any other task(s) that are currently executing. Parsl returns a \textit{future} for each task, allowing the developer to compose dynamic workflows by passing futures between apps.  When the program is executed, Parsl manages task dependencies, data transfers, and execution of tasks on connected resources.

Parsl implements a flexible runtime model via which tasks can be executed on different resources, from laptops to supercomputers. The runtime model builds on two extensible abstractions: the \textit{provider}, a JAAPI that manages the provisioning and management of compute resources (e.g., from batch schedulers or cloud providers);
the \textit{executor} (not to be confused with PSI/J executors) manages the execution of tasks on those resources.
Parsl, like other workflow systems, includes implementations for various providers (e.g., Slurm, PBS, Cobalt, LSF). Parsl has been extended to leverage PSI/J as a community-developed JAAPI. 
Integration was straightforward, as PSI/J implements a compatible model and required making only minor adjustments to certain library calls.

\subsection{RADICAL-Pilot}

RADICAL-Pilot (RP) enables the execution of one or more workloads comprised of heterogeneous tasks on one or more HPC platforms. RP offers five unique features: (1) concurrent execution of tasks with five types of heterogeneity; (2) concurrent execution of multiple workloads on a single pilot, across multiple pilots and across multiple HPC platforms; (3) support of all major DoE and NSF HPC platforms, offering reliable scaling behavior of $O(10^5)$ concurrent tasks on up to $O(6*10^5)$ cores and $O(3.6*10^4)$ GPUs; (4) support for seventeen methods to launch tasks; and (5) integration with third party middleware like workflow and runtime systems. The five types of task heterogeneity supported by RP are: (1) type of task (executable, function or method); (2) parallelism (scalar, MPI, OpenMP, or multiprocess/thread); (3) compute support (CPU and GPU); (4) size (1 hardware thread to 8000 compute nodes); and duration (0 seconds to 48 hours).

As a pilot system, RP schedules tasks concurrently and sequentially, depending on available resources, and defines scheduling policies for executing tasks on all the acquired resources. As such, RP requires scheduling a job on an HPC machine via its batch system to acquire resources, which makes supporting diverse platforms with the same code base challenging. So far, RP has been using RADICAL-SAGA to support all the major batch systems (see Table~\ref{table:jaapis}) but, due to limitations listed in \S\ref{sec:background}, RP has integrated PSI/J to offer the same interoperability but via leaner, more extensible, and easier to integrate interfaces.

\subsection{Swift/T}

Swift/T is a dataflow language with automatic parallelization capabilities~\cite{STC_2014}, with a runtime based on MPI messaging~\cite{Swift_2013}.  Swift/T is designed to run large numbers of small tasks at the CPU core-level granularity at a scale of $O(10^5-10^6)$ processors for tasks that run for $O(10)$ seconds, although larger-granularity tasks, including tasks that are MPI codes, are typical. Swift/T originally used a collection of hand-coded shell scripts to launch itself on scheduler systems including PBS, SLURM, LSF, Cobalt, and plain \texttt{mpiexec} execution.  Using PSI/J provided the opportunity to reduce the maintenance burden of supporting the various schedulers and their often customized installations on exotic large-scale computing systems.

At run time, Swift/T needs to launch its runtime on a certain MPI environment.  The Swift/T runtime, Turbine, is packaged as a Tcl library that links to MPI, ADLB~\cite{ADLB_2010}, and optional application libraries including Python, R, and the JVM library~\cite{Swift_2015}, all at the C level. Running a Swift/T program essentially consists of invoking a command of the form \texttt{mpiexec~tclsh~workflow.tic},
where \texttt{workflow.tic} is a Tcl program translated from the high-level user workflow specification. The complexities involved include setting the job specification and environment as requested by the user with an appropriate request for hardware resources and environment settings for all of the libraries.

A PSI/J module was implemented for Swift/T as done for its other scheduler scripts. However, the PSI/J module is much simpler than those for specific LRMs since all of the complexity is pushed into the PSI/J layer.  A small stub of Python code gathers the settings as encoded by Turbine and presents them to PSI/J, which constructs and launches the job.  Thus, simply by specifying PSI/J as the scheduler, Swift/T users are able to benefit from the portability and feature set of PSI/J without additional work from the Swift/T maintainers.

\subsection{OSPREY: Open Science Platform for Robust Epidemic Analysis}
The COVID-19 pandemic has highlighted both the utility and the difficulties associated with applying computational epidemiology to decision-making in times of crisis and uncertainty. The Open Science Platform for Robust Epidemic Analysis (OSPREY)~\cite{collier_developing_2023} was developed while taking the lessons learned from supporting public health stakeholders during the pandemic. The platform seeks to create a readily deployable capability for producing scenario analyses and forecasts of epidemiological quantities of interest, such as cases, resource needs, and disease outcomes~\cite{Ozik2021population,Hotton2022Impact}, utilizing distributed HPC resources.

One characteristic of epidemic analyses is that their computational demand can vary dramatically over time. Furthermore, computational availability can fluctuate due to demand and resource priorities. Also, for epidemic analyses to be useful as decision support tools, they need to provide actionable insights quickly. To support these requirements, OSPREY uses scalable worker pools, elastic sets of pilot jobs tuned to run specific tasks within the epidemic analyses, such as agent-based disease models or ML/AI computations. These worker pools are dynamically provisioned based on computational demand. The client code that triggers the analyses remotely, e.g., from a researcher laptop, is able to actively monitor the worker pools and terminate them as needed, both of which are done through PSI/J's ability to query and interact with LRMs. The remote authentication and interaction is enabled by Globus Compute~\cite{Chard2020funcX:}. The worker pool is submitted for execution to a LRM using the Swift/T~\cite{wozniak2013swift} framework, which returns a job identifier. Using the PSI/J Python API, a \texttt{psij.Job} object  is created and attach to a native job using an instance of a PSI/J JobExecutor. The Job object's methods and attributes can then be used to query the worker pool job's status or cancel it.

%% file: 06-performance.tex


%

An important characteristic of abstraction libraries is the amount of overhead that they introduce on top of the underlying components that they abstract, since a sufficiently large overhead may be a deterrent in adopting an abstraction library. We present a number of measurements that are meant to quantify the overhead introduced by PSI/J-Python.

We first consider the scaling of the time required to run dummy jobs with the number of jobs using the \textit{local} executor. This measures overheads introduced by the executor itself as well as the overhead of constructing Job objects and querying their status. We run all jobs synchronously and sequentially in loops. We run the experiments on a modern laptop (AMD Ryzen 7 PRO 4750U) running Linux. Results collected for PSI/J-Python, a number of other Python methods of invoking an executable, as well as a Perl version of the same are shown in \figref{scaling-local}. The results show that the local PSI/J executor introduces an overhead of about 10 milliseconds per job. 

All PSI/J executors invoke applications indirectly, using a launcher script, which is specific to the launching mechanism desired by users (e.g., mpirun, srun) but also allows for more complex usage scenarios, such as pre- and post-launch scripts. We first collect results from all methods while uniformly invoking the application through a PSI/J-Python launcher script. In order to ascertain the typical overhead of PSI/J-Python launcher scripts, we compare timing results of the default PSI/J-Python launcher script, with those of a minimal wrapper script which simply runs the job as a subprocess, and, finally, to those without the launcher script. The corresponding results are shown in \figref{scaling-launcher-script}. The results show an average overhead of about 2 milliseconds per job attributable to the use of a launcher script. There is no material difference between the default PSI/J-Python launcher script and a minimal wrapper script.

\begin{figure}[htb]
    \centering
    \includegraphics[width=0.47\textwidth]{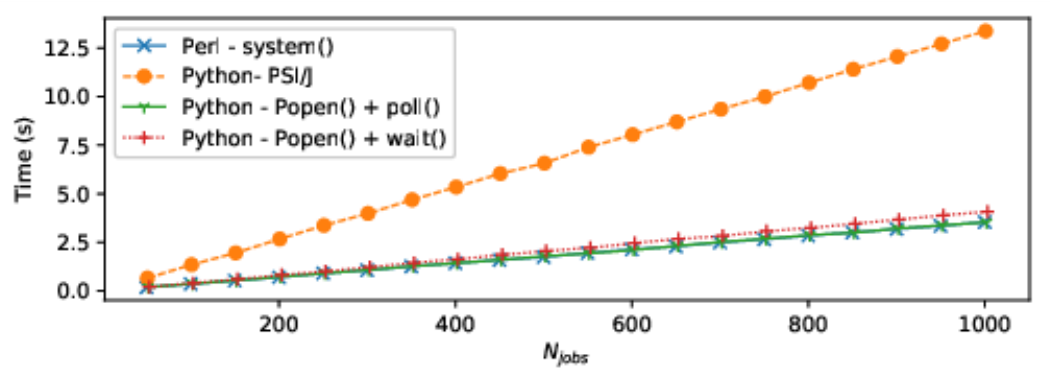}
    \caption{Time required to run $N_{jobs}$ dummy jobs locally using various methods. The ``Popen() + poll()'' method starts the job using the Popen() call and then repeatedly calls ``poll()'' on the Popen object in a busy loop.}
    \figlabel{scaling-local}
\end{figure}

\begin{figure}[htb]
    \centering
    \includegraphics[width=0.47\textwidth]{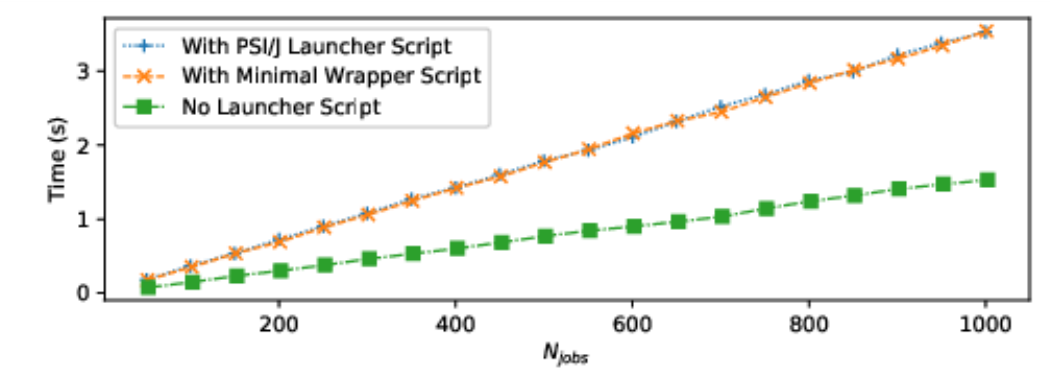}
    \caption{Time required to run $N_{jobs}$ dummy jobs with and without a PSI/J-Python launcher script.}
    \figlabel{scaling-launcher-script}
\end{figure}

Finally, we explore if PSI/J-Python imposes an observable load on a queuing system. We do so by submitting and monitoring a number of jobs while measuring the latency of the queuing system by repeatedly timing a simple \textit{qstat} command that queries the status of a known job. 
We run experiments on three supercomputers at ORNL: Crusher, Frontier, and Summit. These supercomputers use Slurm and LSF schedulers.
The results, presented in \figref{scaling-qstat}, show that, up to 100 jobs, PSI/J does not appear to affect queue performance in a significant way. Nonetheless, a large variance in the time taken to run the \textit{qstat} command is observed.

\begin{figure}[htb]
    \centering
    \includegraphics[width=0.46\textwidth]{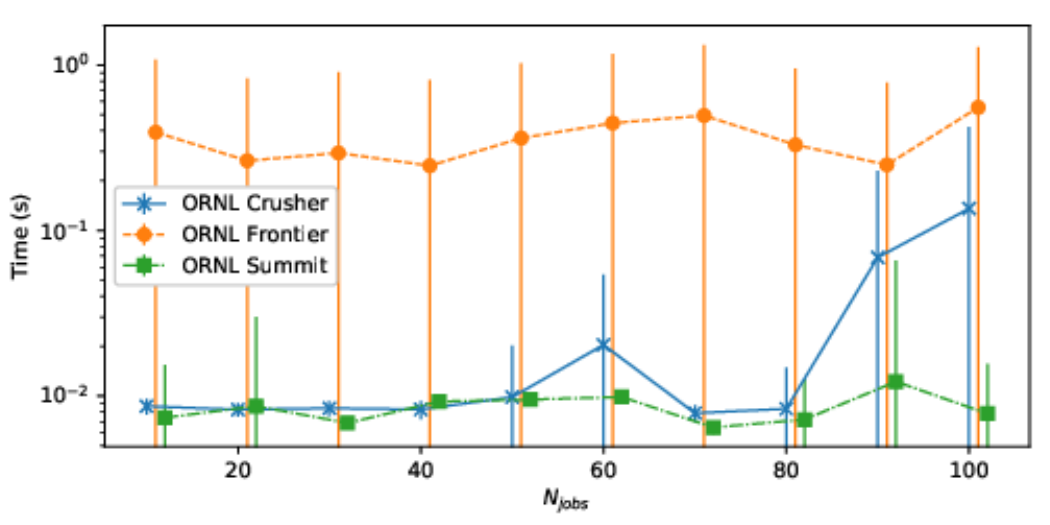}
    \caption{Dependence of the time taken to run a \textit{qstat} command on the number of jobs actively managed by PSI/J.}
    \figlabel{scaling-qstat}
\end{figure}

The numbers shown here suggest that the overheads imposed by the PSI/J Python implementation add an insignificant overhead to the typical jobs and LRMs that the PSI/J Python implementation is expected to manage.

%% file: 07-next.tex

We presented PSI/J, a minimal, composable, and extensible JAAPI specification whose goal is to address many of the hardships associated with achieving a durable JAAPI solution. A reference Python implementation which is meant to encourage community participation is also provided. Additionally, we provide an important tool that allows testing of the reference implementation in ways that would otherwise be significantly challenging.

With most of the JAAPIs described in this paper, it is apparent that the teams responsible for the JAAPIs that ended up getting considerable traction with the community were aware of most challenges and problems involved in designing, implementing, and supporting their solutions successfully. Globus GRAM came at a time when C was the favored language and few modern networking or security libraries or protocols existed. It did what it had to do in order to provide a remote JAAPI solution. The SAGA specification came at a time when the Grid appeared destined for long term domination of the HPC world. It attempted to introduce a standard that addressed the many needs that bridged reality with what was necessary. This gives weight to the idea that extrinsic factors are largely responsible for what the lack of a cohesive JAAPI solution.

We aim to refine both the specification and reference implementation to allow a transition to community-directed support and development in order to mitigate against the likely funding difficulties associated with JAAPIs. We believe that this is a viable solution that is predicated on the assumption that, if provided with a sufficiently well designed JAAPI and a sufficiently well executed reference implementation, the community will stand to gain from supporting a common solution rather than many disparate ones.